\documentclass[draft,aps,prd,amssymb,showpacs]{revtex4}

\def\bea{\begin{eqnarray}}
\def\eea{\end{eqnarray}}
\def\l[{\left[}
\def\r]{\right]}

\begin{document}

\title{One-loop $N$-point equivalence among negative-dimensional,
Mellin-Barnes and Feynman parametrization approaches to Feynman
integrals}

\author{A.T.Suzuki}
\author{E.S.Santos}
\email{suzuki, esdras@ift.unesp.br} \affiliation{Instituto de
F\'isica Te\'orica --- UNESP, R.Pamplona, 145, S\~ao Paulo - SP,
CEP 01405-900, Brazil}

\author{A.G.M.Schmidt}
\email{schmidt@fisica.ufpr.br} \affiliation{Departamento de
F\'{\i}sica --- Universidade Federal do Paran\'a, Caixa Postal
19044, Curitiba - PR, CEP 81531-990, Brazil}


\begin{abstract}

We show that at one-loop order, negative-dimensional,
Mellin-Barnes' (MB) and Feynman parametrization (FP) approaches to
Feynman loop integrals calculations are equivalent. Starting with
a generating functional, for two and then for $N$-point scalar
integrals we show how to reobtain MB results, using
negative-dimensional and FP techniques. The $N-$point result is
valid for different masses, arbitrary exponents of propagators and
dimension.

\end{abstract}
\maketitle

\section{Introduction}

The amazing comparison\cite{laporta} between experimental
determination and theoretical prediction of the anomalous magnetic
momentum of the electron, is the greatest motivation - in our
opinion - to study and develop techniques that allow the precise
calculation of higher order Feynman loop integrals. Recently there
are also interest in studying process like\cite{jatos} $e^+e^-
\rightarrow {\rm 3\; jets}$ and $e^+e^- \rightarrow {\rm 4\;
jets}$ , so loop integrals with five and six external legs must be
known.

Physicists' battle against the tricky Feynman loop integrals is
fought in many fronts. We can cite some of them: integration by
parts method\cite{kotikov} seems to be the most powerful one,
since one can in most cases reduce the number of loops, e.g., the
scalar massless two-loop master can be rewritten as a sum of two
simpler integrals: a two-loop self-energy with an
insertion\cite{halliday,flying} plus the square of a one-loop
self-energy. This is a very simple example -- where one must deal
with a greater number of simpler integrals with powers of
propagators shifted -- of a powerful technique, see for instance a
5-loop calculation in \cite{5-loops}.

Integration by parts is used also associated with another methods.
In fact, one can not evaluate a Feynman loop integral using the
above mentioned technique alone. It simplifies the original
diagram but does not solve it. In order to carry the integration
out Gehrmann and Remiddi \cite{gehrmann} did use the differential
equation method, introduced by Kotikov\cite{eq-dif}, and solved a
large class of difficult of problems. Glover and collaborators
completed the study of the whole class $2\rightarrow 2$ of
two-loop scattering\cite{glover}. Also, Gegenbauer polynomial
method has been used in order to study complicated
process\cite{gegenbauer}.

Other methods that make use of decomposition of complicated
integrals, like the one-loop pentagon\cite{bern-pentagon}, were
developed as well as string inspired ones\cite{string}. See also
\cite{belokurov,bern-gravit,binoth,weinzierl} for other important
approaches, a very powerful numerical technique on
\cite{binoth-num} and for a review on the progress of loop
calculations\cite{tkachov-rev}.

Mellin-Barnes(MB) and negative dimensional integration method
(NDIM) are two other interesting and powerful techniques to tackle
such Feynman integrals. Mellin-Barnes approach relies on the
relation,

\begin{equation}
\left( \sum_{i=1}^{n} z_i\right)^{-B}=\frac{1}{z_1^B (2\pi
i)^{n-1} \Gamma(B)}\int_{-i
\infty}^{i\infty}\Gamma(w_1+w_2+...+w_{n-1}+B)
\prod_{i=1}^{n-1}\left[ dw_i \left(
\frac{z_{i+1}}{z_1}\right)^{w_i}\Gamma(-w_i)\right]. \label{ap2}
\end{equation}
in other words, we rewrite each propagator as a Mellin transform.
However, these parametric integrals are not difficult to solve -
as it happen in the Feynman parametrization approach where the
integrals become more and more complicated - because one can apply
Cauchy theorem and two Barnes' lemmas\cite{boos}. The MB approach
is being greatly used by Tausk\cite{tausk}, Smirnov\cite{smirnov},
Davydychev\cite{davydy2,davydy3} and co-workers in order to tackle
two and three-loop integrals. The results are always expressed as
generalized hypergeometric functions which depend on adimensional
ratios of momenta and/or masses, space-time dimension $D$ and
exponents of propagators.

On the other hand NDIM is a technique whereby it is not necessary
to introduce parametric integrals, the Feynman integral is the
result between the comparison of two calculations: a gaussian-like
integral (the generating functional of negative-dimensional
integrals) and a Taylor expansion of the generating
functional\cite{probing}. It is worth mentioning that in the NDIM
context performing the calculation for a particular set of
exponents of propagators present the same difficulties than
perform the same task for arbitrary values of them. The results
are, just like in the MB approach, given in terms of generalized
hypergeometric functions which depend on the same quantities
mentioned above.

One can then rightfully ask: is there any connection between these
two approaches? The answer is yes. The propose of our paper is to
show that they are equivalent, at least at one-loop order. In
fact, one could argument that the results must be the same if one
correctly applies both methods, we will show this explicitly.
However, it can be useful and interesting to study which of them
is more powerful when the number of external legs increase. In one
hand NDIM demands computer facilities in order to solve a large
number of systems and browse the big number of results; on the
other hand MB does not require computers but the integrals must be
calculated one by one. Another point to observe is that NDIM
relies on grassmannian integrals and MB on Mellin transform, i.e.,
apparently disconnect (as far as we know) subjects. Also, showing
the equivalence between them and knowing the routes which take to
NDIM or to MB one could build, for instance, a technique like NDIM
in order to tackle problems in finite temperature field theory,
like calculations of heat-kernel which can be dealt with using
Mellin integrals.

The outline for our paper is as follows: in section II we present
a step-by-step calculation of 2-point scalar integral starting
with NDIM approach and arriving at an expression originally
obtained by Davydychev\cite{davydy3} using MB approach, then we
repeat the same process using FP. In the next section, we deal
with an arbitrary number $N$ of external legs, also starting in
the negative-dimensional approach and showing how to obtain the
Davydychev's original result calculated in the MB scheme; we carry
out the same the integrals with FP. In the final section we
present our conclusions and a discussion concerning the three
methods.

\section{One-loop 2-point function}

In this section we present the calculations to evaluate the
one-loop two-point scalar integral within NDIM scheme and compare
this result with the obtained by Mellin-Barnes approach. Consider
the integral,
\begin{eqnarray}
I&=&\int d^{D}k \exp\left\{-\alpha
\left[k^2-m_1^2\right]-\beta\left[(k-q)^{2}-m_2^2\right]\right\},
\label{a1}
\end{eqnarray}
which is the usual generating functional for negative-dimensional
integrals. We will always begin with this kind of generating
functionals, for two and for $n$-point scalar integrals, and after
some manipulations arrive at results which were obtained
previously by other authors, using MB approach.

The first step in NDIM context is a series expansion of the above
integral,
\begin{eqnarray}
I&=&\sum^{\infty}_{a_1,a_2=0}\frac{(-\alpha)^{a_1}(-\beta)^{a_2}}{a_1
!a_2 !}J^{(2)}(a_1,a_2;q;m_1,m_2), \label{a2}
\end{eqnarray}
where we define the negative-dimensional integrals,
\begin{eqnarray}
J^{(2)}&=&J^{(2)}(a_1,a_2;q;m_1,m_2)\nonumber \\
&=&\int d^{D}k
{\left[k^2-m_1^2\right]}^{a_1}{\left[(k-q)^{2}-m_2^2\right]}^{a_2}.
\end{eqnarray}
The integration (\ref{a1}) can be easily done,
\begin{eqnarray}
I &=& \left(\frac{\pi}{\alpha+\beta}
\right)^{D/2}\exp\left\{-\left(\frac{\alpha\beta}{\alpha+
\beta}\right)q^2+\alpha m_1^2+\beta m_2^2\right\},
\end{eqnarray}
and the exponential above expanded again in Taylor series,
\begin{eqnarray}
I &=& \left(\frac{\pi}{\alpha+\beta}
\right)^{D/2}\sum_{j_0=0}^{\infty}
\frac{1}{j_0!}\left[-\left(\frac{\alpha\beta}{\alpha+
\beta}\right)q^2+\alpha m_1^2+\beta m_2^2\right]^{j_0}.\nonumber
\end{eqnarray}
Rewrite it as,
\begin{eqnarray}
I &=& \pi^{D/2}\sum_{j_0=0}^{\infty}
\frac{\left(\alpha+\beta\right)^{-D/2+j_0}
\left(m_2^2\right)^{j_0}}{j_0!}\left[1-\frac{\alpha\beta}{
\left(\alpha+\beta\right)^2}\frac{q^2}{m_2^2}-
\frac{\alpha}{\alpha+\beta} \left(1-\frac{m_1^2}{m_2^2}
\right)\right]^{j_0},\label{a3}\end{eqnarray} and then a
multinomial expansion gives us,
\begin{eqnarray}
I = \pi^{D/2} \sum_{j_0,b_1,c_1=0}^{\infty}
\frac{\alpha^{b_1+c_1}\beta^{b_1}\left(\alpha+\beta
\right)^{-D/2+j_0-2b_1-c_1}\left(m_2^2\right)^{j_0}
\Gamma(-j_0+b_1+c_1)}{\Gamma(1+j_0)\Gamma(-j_0)}\frac{\left(\frac{q^2}{m_2^2}\right)^{b_1}}{b_1
!}\frac{\left(1-\frac{m_1^2}{m_2^2}\right)^{c_1}}{c_1 !},
\end{eqnarray}
that using
\begin{eqnarray}
\left[\sum_{i=1}^{n}\alpha_i\right]^{-A} = \frac{1}{\alpha_1^A}
\sum_{j_{1,...n-1}=0}^{\infty}
\frac{(A)_{j_1}}{(1)_{j_1}}\frac{(-j_1)_{j_2}}{(1)_{j_2}}
\frac{(-j_2)_{j_3}}{(1)_{j_3}}...\frac{(-j_{n-2})_{j_{n-1}}}{(1)_{j_{n-1}}}
 \left(-\frac{\alpha_2}{\alpha_1}
\right)^{j_1}\left(-\frac{\alpha_3}{\alpha_2}
\right)^{j_2}\left(-\frac{\alpha_4}{\alpha_3}
\right)^{j_3}...\left(-\frac{\alpha_n}{\alpha_{n-1}}\right)^{j_{n-1}}.
\label{ap1}
\end{eqnarray}
in the factor $(\alpha+\beta)^{-A}$ one obtains,
\begin{eqnarray}
I &=& \pi^{D/2} \sum_{j_0,j_1=0}^{\infty}\sum_{b_1,c_1=0}^{\infty}
\alpha^{-D/2+j_0-b_1-j_1} \beta^{b_1+j_1}
\frac{(-1)^{j_1}(m_2^2)^{j_0}
\Gamma\left(-j_0+b_1+c_1\right)\Gamma\left(D/2-j_0+2b_1+c_1+j_1
\right)}{\Gamma(1+j_0)\Gamma(-j_0)\Gamma
\left(D/2-j_0+2b_1+c_1\right)\gamma(1+j_1)}\nonumber \\
&&\times \left(\frac{q^2}{m_2^2}
\right)^{b_1}\left(1-\frac{m_1^2}{m_2^2}
\right)^{c_1}\frac{1}{b_1!c_1!},\label{a4}
\end{eqnarray}
where $(a)_b = \Gamma(a+b)/\Gamma(a)$, is the Pochhammer symbol.

Compare the power of the parameters $\alpha$ and $\beta$ between
(\ref{a2}) and (\ref{a4}) we have the following constraint
equations,
\begin{eqnarray}
a_1&=&-D/2+j_0-b_1-j_1 \\
a_2&=&b_1+j_1,
\end{eqnarray}
that after solving for $j_0$ and $j_1$ we have
\begin{eqnarray}
j_0&=&a_1+a_2+D/2=\sigma_2\\
j_1&=&a_2-b_1.
\end{eqnarray}
Performing the substitution of this result in (\ref{a4}) and the
analytic continuation to $a_1,a_2\leq 0$, we arrive at,
\begin{eqnarray}
J^{(2)} = \pi^{D/2}(-m_2^2)^{\sigma_2}
\frac{\Gamma(-\sigma_2)}{\Gamma(-a_1-a_2)}
\sum_{b_1,c_1=0}^{\infty} \frac{\left(-\sigma_2\right)_{b_1+c_1}
\left(-a_1\right)_{b_1+c_1}\left(-a_2
\right)_{b_1}}{b_1!c_1!\left(-a_1-a_2\right)_{2b_1+c_1}}
\left(\frac{q^2}{m_2^2} \right)^{b_1}\left(1-\frac{m_1^2}{m_2^2}
\right)^{c_1},\label{a5}\end{eqnarray} which is the well-known
result for 2-point scalar integrals, with different masses, see
\cite{boos}.

\subsection{Two-point function via Feynman Parametrization}\noindent

The most popular technique to deal with loop integrals is
certainly, Feynman parametrization. It is the one the students
learn on field theory courses, and one of the few textbooks
introduce (the other is $\alpha-$parametrization). Depending on
the manipulations one performs, it can turn the original loop
integrals into a hefty one. We will proceed in a slightly
different route. Our aim is to show how one can obtain the
previous results for 2-point functions, given in terms of
hypergeometric functions, using FP since in most cases the results
calculated through FP are written as polylogarithms, $Li_n(z),\;
n=0,1,2,3,4$.

Hypergeometric functions have an advantage over dilogarithms, for
instance, in the case of photon-photon scattering scalar
integrals. The result for $|s/4m^2|<1,\; |t/4m^2|<1$ can be
written as a single Appel function $F_3$ of two variables and 5
parameters, on the other hand, the same result can be recast as a
sum of several $Li_2(z_j)$ functions of complicated arguments
$z_j$, see for instance\cite{photon}.

Consider the function \bea
F^{(2)}&=&F^{(2)}(a_1,a_2;q;m_1;m_2;x_0,x_1)\nonumber \\
&=&\frac{\Gamma(a_1+a_2)}{\Gamma(a_1)\Gamma(a_2)}
\int^{x_0}_{0}dx_1 (x_0-x_1)^{a_1-1}(x_1-x_2)^{a_2-1}  \int
\frac{d^{D}k}{\left\{\left[k^2-m_1^2\right]
\left(x_0-x_1\right)+\left[(k-q)^2-m_2^2\right]
\left(x_1-x_2\right)\right\}^{a_1+a_2}}. \nonumber \label{a6} \eea
where $a_1,a_2\geq 0$. We note that when $x_0=1,\;x_2=0$ we have
the well-known Feynman parametrization to the propagator of
$J^{(2)}$, that is $F^{(2)}=J^{(2)}$. Such modification will turn
simpler the calculation of $N-$point integrals in section IIIA.
This expression can be rewritten as follows, \bea
F^{(2)}&=&\frac{\Gamma(a_1+a_2)(x_0-x_2)^{-D/2}}{\Gamma(a_1)\Gamma(a_2)}
\int^{x_0}_{0}dx_1(x_0-x_1)^{a_1-1}(x_1-x_2)^{a_2-1} \int d^{D}k
\left\{\left[k-\frac{q(x_1-x_2)}{(x_0-x_2)^{1/2}}\right]^2 -m_2^2(x_0- x_2)\right.\nonumber\\
&&\left.
+\frac{q^2(x_0-x_1)(x_1-x_2)}{x_0-x_2}+(m_2^2-m_1^2)(x_0-x_1)\right\}^{-a_1-a_2},
\nonumber \eea that after the evaluation of the integral in $k$
using the well-known formula, \bea \int d^{D}k
\frac{(k^2)^{\alpha}}{\left(k^2+M^2 \right)^{\beta}}
=\pi^{D/2}\left(M^2\right)^{\alpha+D/2-
\beta}\frac{\Gamma(\beta-\alpha-D/2)
\Gamma(\alpha+D/2)}{\Gamma(\beta) \Gamma(D/2)}. \label{ap3} \eea
we have \bea
F^{(2)}&=&\pi^{D/2}\frac{\Gamma(a_1+a_2-D/2)(x_0-x_2)^{-D/2}}{\Gamma(a_1)\Gamma(a_2)}
\int^{x_0}_{0}dx_1(x_0-x_1)^{a_1-1}(x_1-x_2)^{a_2-1}\nonumber \\
&&\times \left\{ -m_2^2(x_0- x_2)
+\frac{q^2(x_0-x_1)(x_1-x_2)}{x_0-x_2}+(m_2^2-m_1^2)(x_0-x_1)\right\}^{D/2-a_1-a_2},
\nonumber \eea Taylor expanding, we get \bea
F^{(2)}&=&\pi^{D/2}(-m_2^2)^{D/2-a_1-a_2}\frac{(x_0-x_2)^{-a_1-a_2}}{\Gamma(a_1)\Gamma(a_2)}
\sum_{b_1=0}^{\infty}\sum^{\infty}_{c_1 =0}(x_0-x_2)^{-b_1 }
x_0^{- b_1- c_1 } \Gamma(a_1+a_2-D/2+b_1+c_1)\nonumber \\
&&\times \frac{1}{b_1 !}\left[\frac{q^2}{m_2^2}\right]^{b_1}
\frac{1}{c_1 !}\left[1-\frac{m_1^2}{m_2^2}\right]^{b_1}
\int^{x_0}_{0}dx_1(x_0-x_1)^{a_1+b_1+c_1-1}(x_1-x_2)^{a_2+b_1-1}.
\nonumber \eea These integrals in $x_1$ can be evaluated, with
$x_0=1,\;x_2=0$, using \bea \prod^{n-2}_{i=0}\left[\int_0^{x_i}
dx_{i+1} (x_i-x_{i+1})^{\alpha_{i+1}-1}\right]
x_{n-1}^{\alpha_n-1}=x_0^{\alpha_1+...+\alpha_n-1}\frac{\prod_{i=1}^{n}
\Gamma(\alpha_i)}{\Gamma(\sum_{i=1}^{n}\alpha_i)}. \label{ap4}
\eea one obtains \bea
F^{(2)}&=&\pi^{D/2}(-m_2^2)^{D/2-a_1-a_2}\frac{\Gamma(a_1+a_2-D/2)}{\Gamma(a_1+a_2)}
\sum_{b_1=0}^{\infty}\sum^{\infty}_{c_1
=0}\frac{(a_1+a_2-D/2)_{b_1+c_1}(a_1)_{b_1+c_1}(a_2)_{b_1}}{b_1!c_1!(a_1+a_2)_{2b_1+c_1}}
\left(\frac{q^2}{m_2^2}\right)^{b_1} \nonumber \\
&&\times \left(1-\frac{m_1^2}{m_2^2}\right)^{c_1},\label{a7} \eea
which is exactly the former result (\ref{a5}). This result show
that, to one-loop two-point, the NDIM, Feynman parametrization and
Mellin-Barnes representation are equivalent. The other kinematical
regions can be obtained through analytic continuation of
hypergeometric function above (see
\cite{boos,triangulos,davydy2}).

\section{N-point function}

In this section we present the generalization of the previous
ideas in order to the obtain the Mellin-Barnes result for the
scalar integral associated to $n$-point function. We consider a
one-loop Feynman diagram with $n$ external legs with momenta:
$p_1=l_2-l_1, p_2=l_3-l_2, ..., p_n=l_1-l_n,$ and internal momenta
$k-l_1, k-l_2,..., k-l_n$. From a similar reasoning, we begin with
the generating functional,

\begin{eqnarray}
I_n&=&\int d^Dk \exp\left\{-\sum_{i=1}^{n}\alpha_i
\left[ \left( k-l_i\right)^2-m_i^2\right]\right\}\label{n-point} \\
&=&\int d^D k\prod_{i=1}^{n}\sum_{a_i=0}^{\infty}
\frac{(-\alpha_i)^{a_i}}{a_i !} \left[
\left(k-l_i\right)^2-m_i^2\right]^{a_i}\nonumber \\
&=&\sum_{a_1,...,a_n =0}^{\infty}\frac{(-\alpha_1)^{a_1}}{a_1
!}... \frac{(-\alpha_n)^{a_n}}{a_n !}
J^{(n)}\left(l_1,m_1,a_1;...;l_n,,m_n,a_n\right),\label{n1}
\end{eqnarray}
where $J^{(n)}\left(l_1,m_1,a_1;...;l_n,,m_n,a_n\right)$
represents the $n-$point functions for negative values of $a_i$
and is given by

\begin{eqnarray}
J^{(n)}&=&J^{(n)}\left(l_1,m_1,a_1;...;l_n,,m_n,a_n\right)\nonumber \\
&=&\int d^D k
\prod_{i=1}^{n}\left[\left(k-l_i\right)^2-m_i^2\right]^{a_i}.
\label{n2}
\end{eqnarray}

The expression (\ref{n1}), after the integration in $k$ can be rewritten of form
\begin{eqnarray}
I_n&=&\left( \frac{\pi}{\sum_{i=1}^{n}\alpha_i}\right)^{D/2} \exp
\left\{ -\frac{\sum_{i>j}^{n}\alpha_i\alpha_j
l^2_{ij}}{\sum_{i=1}^{n}\alpha_i}+\sum_{i=1}^{n} \alpha_i
m_i^2\right\}, \nonumber
\end{eqnarray}
where $l_{ij}=l_i-l_j$. After a new expansion in the right side of
the expression above, we have

\begin{eqnarray}
I_n&=&\left( \frac{\pi}{\sum_{i=1}^{n}\alpha_i}\right)^{D/2}
\sum_{j_0=0}^{\infty}
\frac{1}{\Gamma(1+j_0)}\left[-\frac{\sum_{i>j}^{n}
\alpha_i\alpha_j
l^2_{ij}}{\sum_{i=1}^{n}\alpha_i}+\sum_{i=1}^{n}\alpha_i m_i^2
\right]^{j_0}, \label{nn1}
\end{eqnarray}
using the expansions (\ref{ap2}) and (\ref{ap1}) for $
\left[\sum_{i=1}^{n}\alpha_i\right]$ multinomial, with
$N=n(n-1)/2$ terms, we get

\begin{eqnarray}
I_n&=& \pi^{D/2} \frac{1}{(2\pi
i)^{N+n-1}}\sum_{j_0,...,j_{n-1}=0}^{\infty}
\frac{(-l_{12}^2)^{j_0}}{\Gamma(1+j_0)\Gamma(-j_0)} \int_{-i
\infty}^{i\infty}\frac{(-1)^{j_1}}{\Gamma(1+j_1)}
\prod_{i=2}^{n-1}
\left[\frac{(-1)^{j_i}\Gamma(j_i-j_{i-1})}{\Gamma(1+j_i)
\Gamma(-j_{i-1})}\right]\nonumber \\
&&\times \frac{\Gamma\left({\displaystyle\sum_{j>1, i<j}^{n}}w_{ij} 
+\sum_{i=1}^{n}v_i-j_0\right)\Gamma(D/2+j_0+j_1-\sum_{i=1}^{n}v_i)}{\Gamma
\left(D/2+j_0-\sum_{i=1}^{n}v_i\right)} {\displaystyle\prod_{j>2,
i<j}^{n}}\left[ dw_{ij} \left(
\frac{l_{ij}^2}{l_{12}^2}\right)^{w_{ij}} \Gamma(-w_{ij})\right] \nonumber\\
&&\times \prod_{i=1}^{n} \left[ dv_i
\left(-\frac{m_i^2}{l_{12}^2}\right)^{v_i}
\Gamma(-v_i)\right]\prod_{i}^{n}\left[\alpha_i^{f_i}\right],
\label{n3}\end{eqnarray} where

\begin{eqnarray}
f_1&=&  -j_1-D/2-{\displaystyle\sum_{i\neq 1, i<j}^{n}} w_{ij}+v_1, \nonumber\\
f_2&=&  j_0+j_1-j_2-{\displaystyle\sum_{i\neq 2, i<j}^{n}} w_{ij}-\sum_{i\neq 2}^{n}v_i, \nonumber\\
f_i&=& j_{i-1}-j_i+\sum_{j\neq i}^{n} w_{ij}+v_i,\;\;\;\;i=3,4,...,n-1, \nonumber\\
f_n&=&  j_{n-1}+\sum_{j\neq n}^{n} w_{nj}+v_n. \end{eqnarray}

We need to do now the comparison term-by-term between $\alpha_i$
powers in eq.(\ref{n3}) with the ones of eq.(\ref{n1}). We obtain
$f_i=a_i $ and the solution of system above will be given by,

\begin{eqnarray}
j_0&=&\sum_{i=1}^{n}a_i=\sigma_n, \nonumber\\
j_1&=&  -a_1-D/2-{\displaystyle\sum_{i\neq 1, i<j}^{n}} w_{ij}+v_1,\nonumber\\
j_2-j_1&=&  -a_2+\sigma_n-{\displaystyle\sum_{i\neq 2, i<j}^{n}} w_{ij}-\sum_{i\neq 2}^{n}v_i,\nonumber\\
j_i-j_{i-1}&=& -a_i+\sum_{j\neq i}^{n} w_{ij}+v_i,\;\;\;i=3,4,...,n-1 \nonumber\\
j_{n-1}&=&  a_n-\sum_{j\neq n}^{n} w_{nj}-v_n.
\end{eqnarray}

Performing the substitution of the solutions above in (\ref{n3}),
we arrive at
\begin{eqnarray}
J^{(n)}&=& \pi^{D/2} (l_{12}^2)^{\sigma_n}\frac{(-1)^{a_1+...+a_n}
(1)_{a_1+\epsilon} ...(1)_{a_n+\epsilon}}{(2\pi i)^{N+n-1}
\left[\Gamma(1+\epsilon)\Gamma(-\epsilon)\right]^{n}} \int_{-i
\infty}^{i\infty}{\displaystyle\prod_{j>2, i<j}^{n}} \left[
dw_{ij} \left(
\frac{l_{ij}^2}{l_{12}^2}\right)^{w_{ij}} \Gamma(-w_{ij})\right]\nonumber\\
&&\times \prod_{i=1}^{n} \left[ dv_i
\left(-\frac{m_i^2}{l_{12}^2}\right)^{v_i} \Gamma(-v_i)\right]
\prod_{i=3}^{n}\left[\Gamma\left(-a_i+ \sum_{j\neq
i}^{n}w_{ij}-\sum_{i\neq 2}^{n}v_i\right)
\right]\nonumber \\
&&\times \frac{\Gamma\left({\displaystyle\sum_{j>1,
i<j}^{n}}w_{ij}+
\sum_{i=1}^{n}v_i-\sigma_n\right)\Gamma\left(\sigma_n-a_1-{\displaystyle\sum_{i\neq
1, i<j}^{n}}w_{ij} -\sum_{i=2}^{n}v_i\right)}{\Gamma
\left(D/2+\sigma_n-\sum_{i=1}^{n}v_i \right)}\nonumber \\
&&\times \Gamma\left(-a_2+\sigma_n-{\displaystyle\sum_{i\neq 2,
i<j}^{n}} w_{ij}-\sum_{i \neq 2}^{n}v_i\right), \\
\label{n4}\end{eqnarray} that after carrying out analytic
continuation to negative values of the $a_i$ provides,

\begin{eqnarray}
J^{(n)}&=& \pi^{D/2} (l_{12}^2)^{\sigma_n} \frac{1}{(2\pi
i)^{N+n-1}}\prod_{i=1}^{n}
\left[\frac{1}{\Gamma(-a_i)}\right]\nonumber\\
&&\times \int_{-i \infty}^{i\infty} {\displaystyle\prod_{j>2,
i<j}^{n}}\left[ dw_{ij} \left(
\frac{l_{ij}^2}{l_{12}^2}\right)^{w_{ij}} \Gamma(-w_{ij})\right]
\prod_{i=1}^{n} \left[ dv_i
\left(-\frac{m_i^2}{l_{12}^2}\right)^{v_i}
\Gamma(-v_i)\right]\nonumber \\
&&\times \prod_{i=3}^{n}\left[\Gamma\left(-a_i+\sum_{j\neq
i}^{n}w_{ij} -\sum_{i\neq 2}^{n}v_i\right)\right]
\Gamma\left(-a_2+\sigma_n- {\displaystyle\sum_{i\neq 2, i<j}^{n}}
w_{ij}-\sum_{i\neq 2}^{n}v_i\right)
\nonumber \\
&&\times \frac{\Gamma\left({\displaystyle \sum_{j>1,
i<j}^{n}}w_{ij}+ \sum_{i=1}^{n}v_i- \sigma_n\right)
\Gamma\left(\sigma_n-a_1-{\displaystyle\sum_{i\neq 1,
i<j}^{n}}w_{ij} -\sum_{i=2}^{n}v_i\right)}{\Gamma(D/2+\sigma_n-
\sum_{i=1}^{n}v_i)}. \label{n4-after} \end{eqnarray}

The above result in also an expression for the $n$-point scalar
integrals with arbitrary exponents of propagators and dimension,
in the MB scheme. However, it was not this one the obtained by
Davydychev in \cite{davydy3}. Formula (\ref{n4-after}) is a new
result.

It is important to observe that the above result,
eq.(\ref{n4-after}), is valid since the series is convergent which
means $|l_{ij}^2/l_{12}^2|<1$ and $|m_{i}^2/l_{12}^2|<1$, that is,
external momentum greater than masses. Conversely, Davydychev's
result holds in another kinematical region, namely, where
$|l_{ij}^2/m_{n}^2|<1$ and $|1-m_{i}^2/m_{n}^2|<1$, i.e., when
masses are greater than incoming/outcoming momenta. This result
could be obtained, in principle, from the Davydychev's formula
through analytic continuation. However we stress the point that
such analytic continuation formulas are not known for multiple
hypergeometric series (in general these formulas are known only in
the case of single and double series).

Other form to represent the n-point function can be obtained also
from expansion of (\ref{nn1}), that is
\begin{eqnarray}
I_n&=& \pi^{D/2} \frac{1}{(2\pi
i)^{N+n-1}}\sum_{j_{0,...,n-1}=0}^{\infty}
\frac{(m_n^2)^{j_0}}{\Gamma(1+j_0)\Gamma(-j_0)}
\int_{-i\infty}^{i\infty}\frac{(-1)^{j_1}}{\Gamma(1+j_1)}
\prod_{i=2}^{n-1}
\left[\frac{(-1)^{j_i}\Gamma(j_i-j_{i-1})}{\Gamma(1+j_i)
\Gamma(-j_{i-1})}\right] \nonumber \\
&&\times \frac{\Gamma\left(\sum_{i<j}^{n}w_{ij}+
\sum_{i=1}^{n-1}v_i-j_0\right)\Gamma(D/2+j_1+
{\displaystyle\sum_{i,j\neq n, i<j}^{n}}w_{ij})}{\Gamma
\left(D/2+{\displaystyle\sum_{i,j\neq n, i<j}^{n}}w_{ij} \right)}\nonumber\\
&&\times \prod_{i<j}^{n}\left[ dw_{ij} \left(
-\frac{l_{ij}^2}{m_n^2} \right)^{w_{ij}} \Gamma(-w_{ij})\right]
\prod_{i=1}^{n-1} \left[ dv_i \left(\frac{m_i^2}{m_n^2}
\right)^{v_i}\Gamma(-v_i)\right] \prod_{i}^{n}\left[\alpha_i^{g_i}
\right], \label{n5}\end{eqnarray} where

\begin{eqnarray}
g_i&=&  j_i-j_{i+1}+{\sum_{j\neq i}^{n}} w_{ij}+v_i,\;\;\;i=1,2,...,n-2\nonumber\\
g_{n-1}&=&  j_{n-1}+{\sum_{j\neq n-1}^{n}} w_{n-1,j}+v_{n-1},\nonumber\\
g_n&=&  j_0-j_1-D/2-2\displaystyle\sum_{i,j\neq n, i<j}^{n}
w_{ij}-\sum_{i=1}^{n-1}v_i,
\end{eqnarray}
whose solution, after compare the $\alpha_i$ powers of the
equation (\ref{n5}) with (\ref{n1}), $g_i=a_i$, is given by

\begin{eqnarray}
j_0&=&\sum_{i=1}^{n}a_i=\sigma_n,\nonumber\\
j_1&=& \sigma_n-D/2-2{\displaystyle\sum_{i,j\neq n, i<j}^{n}} w_{ij}-\sum_{i=1}^{n-1}v_i,\nonumber\\
j_{i+1}-j_{i}&=& -a_{i}+{\sum_{j\neq i}^{n}} w_{ij}+v_i,\;\;\;i=1,2,...,n-2\nonumber\\
j_{n-1}&=& a_{n-1}-{\sum_{j\neq n-1}^{n}} w_{n-1,j}-v_{n-1}.
\end{eqnarray}

Analytically continuing and performing the substitution of the
result above in (\ref{n5}) and compare its $\alpha_i$ powers with
(\ref{n1}), we arrive at,
\begin{eqnarray}
J^{(n)}&=& \pi^{D/2}
(-m_n^2)^{\sigma_n}\frac{(-1)^{a_1+...+a_n}(1)_{a_1+\epsilon}...(1)_{a_n+\epsilon}}{(2\pi
i)^{N+n-1}\left[\Gamma(1+\epsilon)\Gamma(-\epsilon)\right]^n}
\int_{-i \infty}^{i\infty}\prod_{i=1}^{n-1} \left[\Gamma(-a_i+
\sum_{j\neq i}^{n}w_{ij}+v_i)\right]\nonumber \\
&&\times \frac{\Gamma\left(\sum_{i<j}^{n}w_{ij}+
\sum_{i=1}^{n-1}v_i-\sigma_n \right)\Gamma(\sigma_n-
{\displaystyle\sum_{i,j\neq n, i<j}^{n}}w_{ij}-
\sum_{i=1}^{n-1}v_i)}{\Gamma\left(D/2+
{\displaystyle\sum_{i,j\neq n, i<j}^{n}}w_{ij} \right)}\nonumber\\
&&\times \prod_{i<j}^{n}\left[ dw_{ij} \left(
-\frac{l_{ij}^2}{m_n^2}\right)^{w_{ij}} \Gamma(-w_{ij})\right]
\prod_{i=1}^{n-1} \left[ dv_i \left(\frac{m_i^2}{m_n^2}
\right)^{v_i} \Gamma(-v_i)\right], \label{n6}
\end{eqnarray}
that can be rewritten of form

\begin{eqnarray}
J^{(n)}&=& \pi^{D/2}
(-m_n^2)^{\sigma_n}\frac{(-1)^{a_1+...+a-n}(1)_{a_1+\epsilon}...(1)_{a_n+\epsilon}}{(2\pi
i)^{N+n-1}\left[\Gamma(1+\epsilon)\Gamma(-\epsilon)\right]^n}
\int_{-i \infty}^{i\infty}\prod_{i<j}^{n}\left[ dw_{ij}
\left( -\frac{l_{ij}^2}{m_n^2}\right)^{w_{ij}} \Gamma(-w_{ij}) \right]\nonumber \\
&&\times \prod_{i=1}^{n-1} \left[ dv_i \left(\frac{m_i^2}{m_n^2}
\right)^{v_i} \Gamma(-v_i)\Gamma(-a_i+\sum_{j\neq
i}^{n}w_{ij}+v_i) \right]\nonumber \\ &&\times
\frac{\Gamma\left(\sum_{i<j}^{n}w_{ij}
+\sum_{i=1}^{n-1}v_i-\sigma_n\right)\Gamma(\sigma_n-
{\displaystyle\sum_{i,j\neq n, i<j}^{n}}w_{ij}
-\sum_{i=1}^{n-1}v_i)}{\Gamma\left(D/2+{\displaystyle\sum_{i,j\neq
n, i<j}^{n}}w_{ij}\right)}, \label{n7}
\end{eqnarray}
that after the analytic continuation to negative values of $a_i$,
we get

\bea J^{(n)}&=&\pi^{D/2} \left(-m_n^2\right)^{\sigma_n}
\frac{\Gamma(-\sigma_n)}{\Gamma(-\sigma_n+D/2)}
\sum_{b_{ij}=0}^{\infty} \sum_{c_l=0}^{\infty}
\frac{\left(-\sigma_n\right)_{\sum_{i>j}b_{ij}+ \sum_{l}c_l}
\prod_{i=1}^{n-1}\left[\left(-a_i\right)_{ \sum_{i\neq j} b_{ij}+
c_i}\right] \left(-a_n\right)_{\sum_{j\neq n} b_{nj}}}{
\left(\sigma_n+D/2\right)_{2 \sum_{i>j}b_{ij}+\sum_{l}c_l}}\nonumber \\
&&\times \prod_{i>j}^{n}\left[\frac{1}{b_{ij}
!}\left(\frac{l^2_{ij}}{m_n^2}
\right)^{b_{ij}}\right]\prod^{n-1}_{l=1}\left[\frac{1}{c_l !}
\left(1-\frac{m_i^2}{m_n^2}\right)^{c_l}\right]\label{n8} \eea 
which was the result obtained by Davydychev in \cite{davydy3}
using MB approach. So, with the generating functional
(\ref{n-point}) as the starting point --- the same which we have,
in a previous paper \cite{n-pontos} in the NDIM approach, used to
show how to obtain a general formula to {\it any} scalar one-loop
Feynman integrals, in covariant gauges --- we were able to
reproduce a MB result eq.(\ref{n8}) and more, to present another
formula (\ref{n4-after}), also valid for $n$-point scalar
integrals.

\subsection{N-Point function via Feynman Parametrization}\noindent

Our final task in this paper is to show how to solve an $N$-point
scalar integral using FP technique. As far as we know there is no
such result in the literature calculated using FP. Of course, it
has to be the same we obtained before using NDIM and Davydychev's
\cite{davydy3} MB approaches.

We start with the function $F^{(n)}$ \bea
F^{(n)}&=&F^{(n)}\left(a_i;l_i;m_i;x_0,x_n\right)\nonumber \\
&=&\frac{\Gamma\left(a_1+...+a_n\right)}{\Gamma
\left(a_1\right)...\Gamma\left(a_n\right)}
\prod^{n-2}_{i=0}\left[\int_0^{x_{i}}dx_i\left(x_{i}-x_{i+1}
\right)^{a_{i+1} -1}\right]\left(x_{n-1}-x_n\right)^{a_n-1}
\nonumber\\ &&\times \int\frac{d^{D}k}{\left\{\prod_{i=1}^{n}\l[
\left(k-l_i\right)^2-m_i^2\r]\left(x_{i-1}-x_i\right)\right\}^{\sum_{i=1}^{n}a_i}},
\eea where $a_i\geq 0$. This function for $x_0=1\;\;x_n=0$,
represent the Feynman parametrization to the integral of type
(\ref{n2}). The integral in $k$ above can be evaluated using
(\ref{ap3}), \bea F^{(n)}&=&\pi^{D/2}\frac{\Gamma(a_1+...+a_n-D/2)
\left(x_0-x_n\right)^{-D/2}}{\Gamma
\left(a_1\right)...\Gamma\left(a_n\right)}
\prod^{n-2}_{i=0}\left[\int_0^{x_{i}}dx_i
\left(x_{i}-x_{i+1}\right)^{a_{i+1} -1}\right]\left(x_{n-1}-x_n
\right)^{a_n-1}\nonumber\\
&&\times \left\{\sum_{i=1}^{n}
\left(l_i^2-m_i^2\right)\left(x_{i-1}-x_i
\right)-\frac{1}{x_0-x_n}\l[
\sum_{i=1}^{n}l_i(x_{i-1}-x_i)\r]^2\right\}^{\sum_{i}a_i-D/2}.
\eea Using now \begin{equation}
\sum_{i=1}^{n}l_i^2\left(x_{i-1}-x_i\right)
-\frac{\l[\sum_{i=1}^{n}l_i(x_{i-1}-x_i) \r]^2}{x_0-x_n} =
\sum_{i>j}^{n} \frac{ (x_{i-1}-x_i)(x_{j-1}-x_j)}{x_0-x_n}
\left(\frac{l_{ij}^2}{m_n^2}\right)\nonumber\\
\label{ap7} \end{equation} and \bea \sum_{i=1}^{n-1}m_i^2
\left(x_{i}-x_{i-1}\right) = -m_n^2(x_0-x_n)+
\sum_{i=1}^{n-1}\left(m_i^2-m_n^2\right)(x_i-x_{i-1}).\nonumber \\
\label{ap8} \eea performing also the Taylor expansion of the
argument of the integral above, we get \begin{eqnarray}
F^{(n)}&=&\pi^{D/2}\left(-m_n^2\right)^{D/2-\sum_i a_i}
\frac{1}{\Gamma\left(a_1\right)...\Gamma\left(a_n\right)}
\sum_{b_{ij}=0}^{\infty}\sum_{c_l=0}^{\infty}
\left(x_0-x_n\right)^{-D/2-2\sum_{i>j}^n b_{ij}-\sum_{l=1}^n
c_l}\nonumber \\ &&\times \Gamma\left(\sum a_i-D/2+\sum_{i>j}
b_{ij}+\sum_{l=1}^{n-1}c_l\right)
\prod_{i>j}^{n}\left[\frac{1}{b_{ij} !}
\left(\frac{l^2_{ij}}{m_n^2}\right)^{b_{ij}}\right]\prod^{n-1}_{l=1}\left[\frac{1}{c_l
!}\left(1-\frac{m_i^2}{m_n^2}\right)^{c_l}\right]\nonumber\\&&\times
\prod^{n-2}_{i=0}\left[\int_0^{x_{i}}dx_i\left(x_{i}-x_{i+1}\right)^{g_{i+1}
-1}\right]\left(x_{n-1}-x_n\right)^{g_n-1},\nonumber\\ \eea where
\bea
g_i&=&a_i+\sum_{j\neq i}b_{ij}+c_i,\;\;\;,i=1,2,...,n-1\\
g_n&=&a_n+\sum_{j\neq n}b_{nj}. \eea The integral above can be
evaluated with help of (\ref{ap4}). If we take $x_0=1,\;x_n=0$,
$F^{(n)}=J^{(n)}$, we arrive \bea
J^{(n)}&=&\pi^{D/2}\left(-m_n^2\right)^{D/2-\sum_i a_i}\frac{\Gamma\left(\sum a_i-D/2
\right)}{\Gamma\left(\sum a_i\right)}\nonumber\\
&&\times \sum_{b_{ij}=0}^{\infty}\sum_{c_l=0}^{\infty}
\frac{\left(\sum a_i -D/2\right)_{\sum_{i>j}b_{ij}+\sum_{l}c_l}
\prod_{i=1}^{n-1}\left[\left(a_i \right)_{\sum_{i\neq j} b_{ij}+
c_i}\right] \left(a_n\right)_{\sum_{j\neq n}
b_{nj}}}{\left(\sigma_n+D/2\right)_{2
\sum_{i>j}b_{ij}+\sum_{l}c_l}}
\prod_{i>j}^{n}\left[\frac{1}{b_{ij}
!}\left(\frac{l^2_{ij}}{m_n^2}\right)^{b_{ij}}\right]\nonumber \\
&&\times \prod^{n-1}_{l=1}\left[\frac{1}{c_l
!}\left(1-\frac{m_i^2}{m_n^2}\right)^{c_l}\right].\label{a13} \eea
This result is the same one obtained in the previous subsection
via NDIM in (\ref{n8}). This agreement show that, in one-loop
level, the NDIM, Feynman parametrization and Mellin-Barnes
representation present the same results and are equivalent: all of
them can be used to solve all scalar Feynman loop integrals at
one-loop order, with general masses, arbitrary exponents of
propagators and dimension.

\section{Discussion and Conclusion}

So far we have made calculations in order to show that the same
class of generating functionals can be used to reproduce MB
results. Depending on which Taylor expansions one carries out one
can proceed in the NDIM or MB routes. The final results will be,
obviously, the same, given in terms of generalized hypergeometric
functions, being the exponents of propagators and space-time
dimension arbitraries.

However, one could ask which of these two routes, if any, is the
one where Feynman integrals become simpler to solve. The first
point to observe what are the tools one has to master in order to
tackle such integrals in both approaches: contour integration,
Cauchy theorem and Barnes' lemmas for MB, and solving system of
algebraic equations for NDIM. So far, so good. Second, the
results, despite they will be the same, have to be worked out
one-by-one in the MB context, on the other hand, using NDIM and
solving the system of algebraic equations gives one all the
possible solutions (generalized hypergeometric functions) for the
Feynman integral in question. Group them is a straightforward
task: linear independent functions have to be summed, each set is
a possible result in a given region if
convergence\cite{triangulos}. Third, the massless case needs to be
known in the case of MB in order to tackle massive integrals; not
so in NDIM.

We can summarize both approaches in following table,

\begin{center}
\begin{tabular}{|l|c|c|}
\hline Step & $MB$ & $NDIM$  \\
\hline 1 & Generating functional & Generating functional
\\ \hline 2 & Solve it & Solve it  \\
\hline  3 & Taylor expand (whole) & Taylor expand (each or whole)
\\ \hline 4 & Mellin transform & Project powers   \\
\hline 5 &  Compare term-by-term & Compare term-by-term
\\
\hline  6 & Solve it for the integral & Solve it for the integral
\\ \hline 7 & Result: parametric integrals & Result: system of algebraic equations \\ \hline  8
& Choose the contour: left or right & Elementary techniques  \\
\hline 9
& Cauchy theorem & Use the results of the systems  \\
\hline 10 & and Barnes' lemmas & Analytically continue to positive $D$  \\
\hline 11 & One have one final result among several & One have all the series (final results)  \\
\hline
\end{tabular}
\end{center}
in the step number 3 one can proceed as we have done in this
paper, expanding the exponential, or as we did in our previous
works taking a Taylor expansion for each argument of the
exponential. The final step, 11, is to be understood in the
following manner: in order to obtain all possible generalized
hypergeometric functions (which come in NDIM) using MB one has to
repeat the above procedure choosing other sequence of contours, we
mean for instance left-left-right-left-right and another one
left-left-right-right-right, these two can give, in principle
distinct generalized hypergeometric series. Some of them, will of
course result in zero, since there can be no poles inside the
contours. These ones are also contained in the NDIM approach,
since some determinants can vanish, a much simpler calculation
that can be implemented in softwares like Mathematica.

The textbook technique, FP, can be made simpler if one introduces
two extra parameters $x_0$ and $x_n$, and takes series expansions
in the parameters $(x_0-x_1),\; (x_1-x_2),\;...,(x_{n-1}-x_{n})$.
In the end of the day one makes $x_0=0$, $x_n=1$ and uses the
well-known beta function integral representation. Then, the
remaining expression is the result written as a generalized
hypergeometric function.

\subsection{Conclusion}

We have shown that negative-dimensional integration method (NDIM),
Feynman parametrization (FP) and Mellin-Barnes' approach to scalar
Feynman loop integrals, at one-loop level, give the same results.
It depends only on how one choose to Taylor expand the generating
functional (\ref{n-point}). We present detailed calculations for
two-point scalar integrals, with arbitrary masses, exponents of
propagators and space-time dimension (in covariant gauges). Then
we tackle a general scalar $N$-point integral, with different
masses, and did show that the general formula of Davydychev
\cite{davydy3} and ours \cite{n-pontos} agree, as well as another
one obtained via FP worked out, as far as we know, for the first
time. It is our opinion however, that NDIM is simpler than MB,
since all the possible results for the integral in question are
obtained simultaneously, and in MB they must be calculated one by
one, or through analytic continuation formulas, if such formulas
were known, depending on the hypergeometric functions. FP is also
a very powerful technique if one introduces two extra parameters
and take Taylor expansions properly. In doing so, FP can become
even simpler than NDIM, since one obtain the full result and does
not have the drawback of searching among a huge amount of possible
solutions.

{\bf Acknowledgments.} ATS and AGMS gratefully acknowledge CNPq
for partial and full financial support, respectively, and ESS
acknowledge CAPES and FAPESP for full financial support.


\begin{thebibliography}{99}


\bibitem{laporta} S.Laporta, E.Remiddi, Phys.Lett.{\bf B356} 390
  (1995); Phys.Lett. {\bf B379} 283 (1996). V.W.Hughes, T.Kinoshita,
  Rev.Mod.Phys. {\bf 71} S133 (1999). S.Laporta, E.Remiddi, Acta
  Phys.Pol. {\bf B28} 959 (1997). P.Mastrolia, E.Remiddi, Nucl. Phys. Proc.Suppl. {\bf 89} (2000) 76.

\bibitem{jatos}
Z.Nagy, Phys.Rev.Lett.{\bf 88} (2002) 122003. S.Weinzierl,
D.A.Kosower, Phys.Rev.{\bf D60} (1999) 054028. M.S. Bilenky,
G.Rodrigo, A.Santamaria, Nucl. Phys. Proc. Suppl. {\bf 74} (1999)
53. W.Bernreuther, A.Brandenburg, P.Uwer, Nucl. Phys. Proc. Suppl.
{\bf 96} (2001) 79.  L.W. Garland, T. Gehrmann, E.W.N. Glover, A.
Koukoutsakis, E. Remiddi, Nucl. Phys. {\bf B642} (2002) 227.

\bibitem{kotikov}  F.V.Tkachov, Phys.Lett.{\bf B100} (1980) 65.
K.G.Chetyrkin, F.V.Tkachov, Nucl.Phys. {\bf B 192} (1981) 159. An
interesting application of the method can be found in: C.
Anastasiou, T. Gehrmann, C. Oleari, E. Remiddi, J.B. Tausk,
Nucl.Phys.{\bf B580} (2000) 577.

\bibitem{halliday} I.G.Halliday, R.M.Riccota, Phys.Lett. {\bf B193}
  241 (1987).  G.V.Dunne, I.G.Halliday, Phys.Lett. {\bf B193} 247
  (1987); Nucl.Phys. {\bf B308} 589 (1988). D.J.Broadhurst, Phys.Lett.
  {\bf B197} 179 (1987).

\bibitem{flying} A.T.Suzuki, A.G.M.Schmidt, Eur.Phys.J.{\bf C5} (1998) 175.

\bibitem{5-loops} S.G. Gorishnii, S.A. Larin, F.V. Tkachov, K.G. Chetyrkin,
Phys.Lett.{\bf B132} (1983) 351.

\bibitem{gehrmann} T. Gehrmann, E. Remiddi, Nucl.Phys.{\bf B601} (2001)
287. {\it ibid}, Nucl.Phys.Proc.Suppl.{\bf 89} (2000) 251.

\bibitem{eq-dif} A.V. Kotikov, Phys.Lett. {\bf B254} (1991) 158;
Phys.Lett. {\bf B259} (1991) 314; Phys.Lett. {\bf B267} (1991)
123. The last paper also deals with $N$-point integrals.

\bibitem{glover} C. Anastasiou, E.W.N. Glover, C. Oleari, M.E.
Tejeda-Yeomans, Nucl.Phys. {\bf B601} (2001) 318; Nucl.Phys. {\bf
B605} (2001) 486. E.W.N. Glover, C. Oleari, M.E. Tejeda-Yeomans,
Nucl.Phys.{\bf B605} (2001) 467. V. del Duca, E.W.N. Glover, JHEP
{\bf 0110} (2001) 035.

\bibitem{gegenbauer}
K.G.Chetyrkin, A.L.Kataev, F.V.Tkachov, Nucl. Phys. {\bf B 174}
(1980) 345. W. Celmaster,  R.J. Gonsalves, Phys.Rev. {\bf D21}
(1980) 3112. A. Terrano, Phys.Lett. {\bf B93} (1980) 424. B.
Lampe, G. Kramer, Phys.Scr. {\bf 28} (1983) 585. A.V. Kotikov,
Phys.Lett. {\bf B375} (1996) 240.

\bibitem{bern-pentagon} Z.Bern, L.J.Dixon, D.A.Kosower, Nucl.Phys.{\bf B412} (1994) 751.

\bibitem{string} H-T.Sato, M.G.Schmidt, C.Zahlten, Nucl.Phys.{\bf B579} (2000)
492. H-T.Sato, M.G.Schmidt, Nucl.Phys.{\bf B560} (1999) 551.

\bibitem{belokurov}
V.V.Belokurov, N.I.Usyukina, J.Phys.{\bf A16} (1983) 2811.

\bibitem{bern-gravit} Z.Bern, L.Dixon, C.Schmidt, Phys.Rev.{\bf D66} (2002)
074018. Z.Bern, Living Rev.Rel.{\bf 5} (2002) 5. Z.Bern, A. De
Freitas, L.Dixon, H.L.Wong, Phys.Rev.{\bf D66} (2002) 085002.

\bibitem{binoth}
T. Binoth, J.P. Guillet, G. Heinrich, C. Schubert, Nucl.Phys.{\bf
B615} (2001) 385. G. Heinrich, T. Binoth,
Nucl.Phys.Proc.Suppl.{\bf 89} (2000) 246. T. Binoth, J.P. Guillet,
G. Heinrich, Nucl.Phys.{\bf B572} (2000) 361.

\bibitem{weinzierl}
S.Weinzierl, J.Phys.{\bf G26} (2000) 654. S.Weinzierl,
D.A.Kosower, Phys.Rev.{\bf D60} (1999) 054028. A.I.Davydychev,
P.Osland, O.V.Tarasov, Phys.Rev.{\bf D54} (1996) 4087;
Erratum-ibid.{\bf D59} (1999) 109901.

\bibitem{binoth-num}
T. Binoth, G. Heinrich, N. Kauer, Nucl.Phys.{\bf B654} (2003),
277. T. Binoth,G. Heinrich, hep-ph/0305234.

\bibitem{tkachov-rev} F.V.Tkachov, Nucl.Instrum.Meth.{\bf A389} (1997)
309. Z.Bern, preprint hep-ph/0212406.

\bibitem{boos} \'E. \'E. Boos and A. I. Davydychev, Theor. Math. Phys. {\bf 89} (1991) 1052.

\bibitem{tausk} J.B.Tausk, Phys.Lett.{\bf B469} (1999) 225. C.Anastasiou, J.B. Tausk,
M.E. Tejeda-Yeomans, Nucl. Phys. Proc. Suppl.{\bf 89} (2000) 262.

\bibitem{smirnov} V.A. Smirnov, Phys.Lett.{\bf B547} (2002) 239;
hep-ph/0305142.

\bibitem{davydy2}  A. I. Davydychev, J. Phys.{\bf A 25} (1992) 5587.

\bibitem{davydy3} A. I. Davydychev, J. Math. Phys.{\bf 33} (1992) 358.

\bibitem{probing} A.T.Suzuki, A.G.M.Schmidt, R.Bent\'{\i}n,
Nucl.Phys. {\bf B537} (1999) 549.

\bibitem{photon} A.I.Davydychev, {\it Zvenigorod Quarks 1992:260-270}; hep-ph/9307323.
A.T.Suzuki, A.G.M.Schmidt, J.Phys.A{\bf 31} (1998) 8023.

\bibitem{triangulos} A.T.Suzuki, E.S.Santos, A.G.M.Schmidt, Eur.Phys.J. {\bf C26}
(2002) 125. See also the preprint hep-ph/0210148.

\bibitem{n-pontos} A.T.Suzuki, E.S.Santos, A.G.M.Schmidt, hep-ph/0210083.



\end{thebibliography}
\end{document}